\documentclass[a4paper,fleqn]{article}
\usepackage{amsmath}

\begin{document}

\title{\textbf{An integrable hierarchy without a recursion operator}}

\author{\textsc{Sergei Sakovich}\bigskip \\
\small Institute of Physics, National Academy of Sciences of Belarus \\
\small sergsako@gmail.com}

\date{}

\maketitle

\begin{abstract}
We study the Lax integrability of a nonlinear system of two coupled second-order evolution equations introduced by Ibragimov and Shabat. For this system we find a zero-curvature representation with an essential parameter, construct an infinite integrable hierarchy which the system belongs to, and show that this hierarchy does not possess a recursion operator.
\end{abstract}

\section{Introduction}

Despite its simplicity, the following nonlinear system of two coupled second-order evolution equations
\begin{equation}
u_t = u_{xx} + \frac{1}{2} v^2 , \qquad v_t = 2 v_{xx} , \label{e1}
\end{equation}
introduced in \cite{IS}, has a rich and interesting history.

For a long time, this system was thought to possess only one local generalized symmetry, namely, a third-order one. See, for instance, the original paper \cite{IS} itself and the first edition of the book \cite{Olv}. However, it was pointed out later, in Exercise 5.16(a) of the second edition of \cite{Olv}, that the nonlinear system \eqref{e1} does possess local generalized symmetries of higher orders, and it was proposed there to find a higher order symmetry and a recursion operator for this system. Nevertheless, no recursion operator of \eqref{e1} appeared in the literature.

According to the classification made in \cite{BSW}, the system \eqref{e1} belongs to one of the nine exceptional cases of Bakirov-type systems, as the $\mathcal{B}_2 [ \frac{1}{2} ]$ case, therefore it possesses infinitely many local generalized symmetries. Moreover, a recurrent procedure was given in \cite{BSW}, which allows to construct a symmetry of order $n$ for the system \eqref{e1} from its two symmetries of orders $n-1$ and $n-2$. Such a recurrent procedure is, however, not a recursion operator, in the sense that a recursion operator should produce one symmetry from one symmetry.

It is easy to find a formal recursion operator $R$ of the system \eqref{e1}, or a formal symmetry, which satisfies the condition $D_t (R) = [ F , R ]$, the result being
\begin{gather}
R =
\begin{pmatrix}
D_x & d \\ 0 & 2 D_x
\end{pmatrix}
, \label{e2} \\
d = v D_x^{-1} + v_x D_x^{-2} + v_{xx} D_x^{-3} + v_{xxx} D_x^{-4} + \dotsb , \label{e3}
\end{gather}
where $D_t$ and $D_x$ stand for the total derivatives, the square brackets denote the commutator, and $F$ is the Fr\'echet derivative of the right-hand side of the evolutionary system \eqref{e1}. According to \cite{Ser}, such formal recursion operators do exist for wide classes of block-triangular evolutionary systems. It is not clear, however, how to apply the expansion \eqref{e3} to local expressions, or how to bring the formal recursion operator \eqref{e2} with \eqref{e3} into a closed form. We do not believe that the Ibragimov--Shabat system \eqref{e1} possesses any recursion operator of the usual quotient form
\begin{equation}
R = M N^{-1} , \label{e4}
\end{equation}
where $M$ and $N$ are linear matrix differential operators with local coefficients.

All that was about generalized symmetries. In the present paper, we consider the Ibragimov--Shabat system \eqref{e1} from a different point of view. Namely, we study its integrability in the Lax sense. In Section~\ref{s2}, we find a zero-curvature representation of the system \eqref{e1}, with an essential (spectral) parameter. In Section~\ref{s3}, we give a gauge-invariant description of the obtained linear spectral problem. In Section~\ref{s4}, we construct an infinite integrable hierarchy which the system \eqref{e1} belongs to. In Section~\ref{s5}, we show that this hierarchy has no recursion operator. Section~\ref{s6} contains concluding remarks. We use the cyclic basis method \cite{S95,S04} to study the spectral problem and to obtain the hierarchy. The paper is partially based on a section of our preprint \cite{S03a}.

\section{Zero-curvature representation} \label{s2}

Let us find a Lax pair of the Ibragimov--Shabat system \eqref{e1}, in the form of a zero-curvature representation (ZCR).

A ZCR is the compatibility condition
\begin{equation}
Z \equiv D_t X - D_x T + [ X , T ] = 0 \label{e5}
\end{equation}
of the overdetermined linear system
\begin{gather}
\Psi_x = X \Psi , \label{e6} \\
\Psi_t = T \Psi , \label{e7}
\end{gather}
where $X$ and $T$ are $(m \times m)$-dimensional matrix functions of $u$, $v$ and their finite-order derivatives with respect to $x$ (since $t$-derivatives can be expressed in terms of $x$-derivatives via \eqref{e1}), $\Psi (x,t)$ is a $m$-component column, and the condition \eqref{e5} is satisfied by any solution  $u(x,t)$ and $v(x,t)$ of \eqref{e1}.

Any linear transformation
\begin{equation}
\Psi' = G \Psi , \qquad \det G \ne 0 , \label{e8}
\end{equation}
where $G$ is a $(m \times m)$-dimensional matrix function of $u$, $v$ and their finite-order $x$-derivatives, generates the gauge transformation
\begin{gather}
X' = G X G^{-1} + ( D_x G ) G^{-1}, \label{e9} \\
T' = G T G^{-1} + ( D_t G ) G^{-1}, \label{e10}
\end{gather}
which leads to the equivalence transformation $Z' = G Z G^{-1}$ of the ZCR. This makes sense to study ZCRs by gauge-invariant methods.

Under the simplest assumption that $X = X(u,v)$ and $T = T(u, u_x , v, v_x)$, we get by direct analysis of \eqref{e5} with \eqref{e1} the following expressions:
\begin{gather}
X = U u + V v + W , \label{e11} \\
T = U u_x + 2 V v_x - [U,W] u - 2 [V,W] v + Y , \label{e12}
\end{gather}
where $U$, $V$, $W$ and $Y$ are constant $(m \times m)$-dimensional matrices (traceless, without loss of generality) which satisfy the system of commutator relations
\begin{gather}
[U,[U,W]] = 0 , \qquad [U,[V,W]] = 0 , \qquad [V,[V,W]] = \frac{1}{4} U , \notag \\
[W,[U,W]] = [U,Y] , \qquad [W,[V,W]] = \frac{1}{2} [V,Y] , \notag \\
[W,Y] = 0 , \qquad [U,V] = 0 . \label{e13}
\end{gather}

In our experience, the only effective method to solve systems like \eqref{e13} is a brute-force attack, as we did, e.g., in \cite{S99,KSY,S11,SS,S18}. We successively take the matrix $V$ in all possible Jordan forms, use a computer algebra program to solve the system of polynomial equations \eqref{e13} for the unknown elements of the matrices involved, suppress the excessive arbitrariness of solutions by gauge transformations \eqref{e9} and \eqref{e10} with constant $G$, and increase the matrix dimension $m$ if no nontrivial solution appeared. In this way, we obtain a nontrivial solution at $m=4$, which leads via \eqref{e11} and \eqref{e12} to the following expressions:
\begin{gather}
X =
\begin{pmatrix}
\lambda & v & 0 & - 8 u \\
0 & 0 & 1 & 0 \\
0 & 0 & 0 & v \\
0 & 0 & 0 & - \lambda
\end{pmatrix}
, \label{e14} \\
T =
\begin{pmatrix}
2 \lambda^2 & 2 \lambda v + 2 v_x & - 2 v & - 16 \lambda u - 8 u_x \\
0 & 0 & 4 \lambda & 2 v \\
0 & 0 & 0 & 2 \lambda v + 2 v_x \\
0 & 0 & 0 & - 2 \lambda^2
\end{pmatrix}
, \label{e15}
\end{gather}
where $\lambda$ is a parameter. If needed, expressions for $U$, $V$, $W$ and $Y$ are easily seen from \eqref{e14} and \eqref{e15}.

We have found the ZCR \eqref{e5}, or the Lax pair \eqref{e6} and \eqref{e7}, with the matrices $X$ and $T$ given by \eqref{e14} and \eqref{e15}, for the Ibragimov--Shabat system \eqref{e1}.

\section{Gauge-invariant description} \label{s3}

A gauge-invariant description of ZCRs appeared for the first time in \cite{Mar}, in a very general and abstract form, applicable to any non-overdetermined systems of partial differential equations. Later and independently, another gauge-invariant description of ZCRs, applicable to systems of evolution equations only, appeared in \cite{S95}, where the concept of cyclic bases of ZCRs was introduced. In the present paper, we use the cyclic basis method and follow the terminology of \cite{S95,S04}.

The matrix $X$ \eqref{e14} of the ZCR \eqref{e5}, obtained for the Ibragimov--Shabat system, contains no derivatives of $u$ and $v$. Therefore the characteristic matrices of this ZCR are simply
\begin{equation}
C_u = \frac{\partial X}{\partial u} , \qquad C_v = \frac{\partial X}{\partial v} , \label{e16}
\end{equation}
and the characteristic form of this ZCR is
\begin{equation}
u_t C_u + v_t C_v = \nabla T , \label{e17}
\end{equation}
where the covariant derivative operator $\nabla$ is defined as $\nabla (\cdot) = D_x (\cdot) - [ X , (\cdot) ]$. We repeatedly apply the operator $\nabla$ to the characteristic matrices $ C_u$ and $C_v$, and find in this way that the cyclic basis is three-dimensional, consists of the matrices $ C_u$, $C_v$ and $\nabla C_v$, and has the closure equations
\begin{equation}
\nabla C_u = - 2 \lambda C_u , \qquad \nabla^2 C_v = - \frac{1}{4} v C_u - \lambda^2 C_v - 2 \lambda \nabla C_v . \label{e18}
\end{equation}

The dimensions of cyclic bases of ZCRs and the coefficients of closure equations are gauge invariants. It is very convenient to use these invariants to verify whether a parameter in a given Lax pair is an essential (spectral) parameter or this parameter can be removed (gauged out) by a gauge transformation, and whether two given Lax pairs can (or cannot) be related to each other by a gauge transformation \cite{S95,S04,S03a,S99,KSY,S11,SS,S18,S94,S20,S03,S14}. In the present case, we easily see from the coefficients of the closure equations \eqref{e18} that $\lambda$ is an essential parameter.

The three matrices $ C_u$, $C_v$ and $\nabla C_v$ of the cyclic basis, however, are not sufficient to decompose the matrix $T$ \eqref{e15} over them. Therefore we have to use also the singular basis, that is, such matrices $S$ that $\nabla S$ can be decomposed over the cyclic basis but $S$ themselves cannot. We find two,
\begin{equation}
S_1 =
\begin{pmatrix}
1 & 0 & 0 & 0 \\
0 & 0 & 0 & 0 \\
0 & 0 & 0 & 0 \\
0 & 0 & 0 & - 1
\end{pmatrix}
, \qquad
S_2 =
\begin{pmatrix}
0 & 0 & 0 & 0 \\
0 & 0 & 1 & 0 \\
0 & 0 & 0 & 0 \\
0 & 0 & 0 & 0
\end{pmatrix}
, \label{e19}
\end{equation}
with the closure equations
\begin{equation}
\nabla S_1 = 2 u C_u + v C_v , \qquad \nabla S_2 = - \lambda v C_v - v \nabla C_v . \label{e20}
\end{equation}
Now, the matrix $T$ \eqref{e15} can be decomposed as
\begin{equation}
T = ( u_x + 2 \lambda u ) C_u + 2 v_x C_v - 2 v \nabla C_v + 2 \lambda^2 S_1 + 4 \lambda S_2 . \label{e21}
\end{equation}

The expressions \eqref{e17}, \eqref{e21}, \eqref{e18} and \eqref{e20} represent the Ibragimov--Shabat system \eqref{e1} in a gauge-invariant way.

\section{Infinite integrable hierarchy} \label{s4}

Nonlinear equations, integrable in the Lax sense, appear usually as members of their respective hierarchies. Within a hierarchy, Lax pairs of all members have one and the same $x$-part (a spectral problem), whereas $t$-parts (time evolutions) differ. Also usually, the members of an integrable hierarchy are related to each other by integer degrees of a recursion operator.

If a given matrix $X$ of the linear problem \eqref{e6} contains no essential parameter (parameters are absent or can be gauged out), then there is always a continual class (not a habitual discrete hierarchy) of evolution equations which all possess ZCRs \eqref{e5} with this matrix $X$, and most of those equations are non-integrable, of course \cite{S95,S04,S05a}. Moreover, continual classes can also appear for linear problems with essential parameters, and this indicates that two dependent variables in such a spectral problem can be merged into one new dependent variable by a gauge transformation \cite{S95,S05}. Finally, if a given spectral problem really leads to a discrete hierarchy, the cyclic basis technique can also be useful to derive a recursion operator of this hierarchy \cite{S95,S04,S03a,SS,S03,KKS}.

Let us find all evolutionary systems
\begin{gather}
u_t = f ( u, v, \dotsc , u_{x \dotsc x} , v_{x \dotsc x} ) , \notag \\
v_t = g ( u, v, \dotsc , u_{x \dotsc x} , v_{x \dotsc x} )  \label{e22}
\end{gather}
which possess ZCRs \eqref{e5} with the matrix $X$ given by \eqref{e14}. Then the matrices $T$ necessarily have the form
\begin{equation}
T = p C_u + q C_v + r \nabla C_v + \sigma S_1 + \tau S_2 , \label{e23}
\end{equation}
where $p$, $q$, $r$, $\sigma$ and $\tau$ are functions of $u$, $v$ and their finite-order $x$-derivatives. The characteristic form \eqref{e17} of these ZCRs, with $u_t$ and $v_t$ given by \eqref{e22} and $T$ given by \eqref{e23}, leads us via the closure equations \eqref{e18} and \eqref{e20} to the following relations:
\begin{gather}
D_x \sigma = D_x \tau = 0, \label{e24} \\
q = - D_x r + 2 \lambda r + \tau v , \label{e25}
\end{gather}
and
\begin{gather}
f = D_x p - 2 \lambda p - \frac{1}{4} v r + 2 \sigma u , \notag \\
g = - D_x^2 r + 2 \lambda D_x r - \lambda^2 r + \sigma v + \tau ( v_x - \lambda v ) . \label{e26}
\end{gather}
The conditions \eqref{e24} show that $\sigma$ and $\tau$ are constants, in the sense that they do not depend on $u$, $v$ and derivatives of $u$ and $v$. The relation \eqref{e25} just expresses $q$ via $r$ and $\tau$, and it is used to construct the corresponding matrices $T$ \eqref{e23}. The central role belongs to the expressions \eqref{e26} which determine all the represented systems \eqref{e22}, provided that the conditions
\begin{equation}
\frac{\partial f}{\partial \lambda} = \frac{\partial g}{\partial \lambda} = 0 \label{e27}
\end{equation}
are satisfied, while the functions $p$ and $r$ and the constants $\sigma$ and $\tau$ can (and do) depend on $\lambda$. In their turn, these conditions \eqref{e27} determine all the admissible functions $p$ and $r$ and constants $\sigma$ and $\tau$.

In order not to deal with infinite-order derivatives and nonlocal variables related to $u$ and $v$, we consider $p$, $r$, $\sigma$ and $\tau$ as polynomials in $\lambda$:
\begin{gather}
p = \sum_{i=0}^n p_i \lambda^i , \qquad r = \sum_{i=0}^n r_i \lambda^i , \notag \\
\sigma = \sum_{i=0}^n \sigma_i \lambda^i , \qquad \tau = \sum_{i=0}^n \tau_i \lambda^i , \label{e28}
\end{gather}
where the functions $p_i$ and $r_i$ (of $u$, $v$ and their finite-order $x$-derivatives) and the constants $\sigma_i$ and $\tau_i$ do not depend on $\lambda$. We substitute the expansions \eqref{e28} to the relations \eqref{e26}, collect terms with $\lambda^k$, separately for $k = 0, 1, 2, \dotsc , n+2$, and obtain in this way the expressions for $p_i$, $r_i$, $\sigma_i$ and $\tau_i$, as well as for $f$ and $g$ which appear at $\lambda^0$ due to \eqref{e27}. For low orders $n$, we get the following results.

For $n=0$, we get $p = 0$, $r = 0$, $\sigma = \sigma_0$, $\tau = 0$, and
\begin{equation}
\begin{pmatrix}
f \\
g
\end{pmatrix}
= \sigma_0
\begin{pmatrix}
2u \\
v
\end{pmatrix}
,
\label{e29}
\end{equation}
where $\sigma_0$ is an arbitrary constant.

For $n=1$, we get $p = \sigma_1 u$, $r = 0$, $\sigma = \sigma_0 + \sigma_1 \lambda$, $\tau = \sigma_1$, and
\begin{equation}
\begin{pmatrix}
f \\
g
\end{pmatrix}
= \sigma_1
\begin{pmatrix}
u_x \\
v_x
\end{pmatrix}
+ \sigma_0
\begin{pmatrix}
2u \\
v
\end{pmatrix}
,
\label{e30}
\end{equation}
where $\sigma_0$ and $\sigma_1$ are arbitrary constants.

For $n=2$, we get $p = \frac{1}{2} \sigma_2 u_x + \sigma_1 u + \sigma_2 u \lambda$, $r = - \sigma_2 v$, $\sigma = \sigma_0 + \sigma_1 \lambda + \sigma_2 \lambda^2$, $\tau = \sigma_1 +2 \sigma_2 \lambda$, and
\begin{equation}
\begin{pmatrix}
f \\
g
\end{pmatrix}
= \sigma_2
\begin{pmatrix}
\frac{1}{2} u_{xx} + \frac{1}{4} v^2 \\
v_{xx}
\end{pmatrix}
+ \sigma_1
\begin{pmatrix}
u_x \\
v_x
\end{pmatrix}
+ \sigma_0
\begin{pmatrix}
2u \\
v
\end{pmatrix}
,
\label{e31}
\end{equation}
where $\sigma_0$, $\sigma_1$ and $\sigma_2$ are arbitrary constants. The Ibragimov--Shabat system \eqref{e1} appears in \eqref{e31} at $\sigma_2 = 2$ and $\sigma_1 = \sigma_0 = 0$.

For $n=3$, we get $p = \frac{1}{4} \sigma_3 ( u_{xx} + v^2 ) + \frac{1}{2} \sigma_2 u_x + \sigma_1 u + \left( \frac{1}{2} \sigma_3 u_x + \sigma_2 u \right) \lambda + \sigma_3 u \lambda^2$, $r = - \sigma_3 v_x - \sigma_2 v - 2 \sigma_3 v \lambda$, $\sigma = \sigma_0 + \sigma_1 \lambda + \sigma_2 \lambda^2 + \sigma_3 \lambda^3$, $\tau = \sigma_1 +2 \sigma_2 \lambda + 3 \sigma_3 \lambda^2$, and
\begin{gather}
\begin{pmatrix}
f \\
g
\end{pmatrix}
= \sigma_3
\begin{pmatrix}
\frac{1}{4} u_{xxx} + \frac{3}{4} v v_x \\
v_{xxx}
\end{pmatrix}
+ \sigma_2
\begin{pmatrix}
\frac{1}{2} u_{xx} + \frac{1}{4} v^2 \\
v_{xx}
\end{pmatrix}
\notag \\
\qquad \qquad \qquad
+ \sigma_1
\begin{pmatrix}
u_x \\
v_x
\end{pmatrix}
+ \sigma_0
\begin{pmatrix}
2u \\
v
\end{pmatrix}
,
\label{e32}
\end{gather}
where $\sigma_0$, $\sigma_1$, $\sigma_2$ and $\sigma_3$ are arbitrary constants.

The expressions \eqref{e29}--\eqref{e32} demonstrate a typical discrete hierarchy structure. The right-hand sides
\begin{equation}
h =
\begin{pmatrix}
f \\
g
\end{pmatrix}
\label{e33}
\end{equation}
of the represented systems \eqref{e22} are the linear superpositions
\begin{equation}
h = \sum_{i=0}^n \sigma_i h^{(i)} \label{e34}
\end{equation}
of the right-hand sides $h^{(i)}$ of $i$th-order members of the hierarchy, with arbitrary constant coefficients $\sigma_i$, whereas $h^{(i)}$ themselves contain no parameters. The origin of the superpositions \eqref{e34} is, of course,  the linearity of the relations \eqref{e26} with respect to the sets $(p, r, \sigma, \tau)$. However, we still have to prove that the hierarchy is infinite, in the sense that local expressions $h^{(n)}$ exist for arbitrarily high orders $n$.

To obtain the higher-order expressions $h^{(n)}$, we consider the expansions \eqref{e28} with $\sigma_0 = \sigma_1 = \dotsc = \sigma_{n-1} = 0$ and $\tau_0 = \tau_1 = \dotsc = \tau_{n-2} = 0$. Note that we retain the possibility of nonzero $\tau_{n-1}$, because we have already found that $\tau_n = 0$ and $\tau_{n-1} = n \sigma_n$ for $n = 0, 1, 2, 3$ (and we have to prove the same for higher orders $n$).

For a sufficiently high order $n$, the terms in \eqref{e26} with $\lambda^k$, successively for $k = 0, 1, \dotsc , n-2$ (that is, where $\sigma$ and $\tau$ have no effect), give us the following expressions:
\begin{gather}
h^{(n)} = M s_0 , \label{e35} \\
M s_1 + L s_0 = 0 , \label{e36} \\
M s_k + L s_{k-1} + K s_{k-2} = 0 , \qquad k = 2 , \dotsc , n-2 , \label{e37}
\end{gather}
with the notations
\begin{gather}
M =
\begin{pmatrix}
D_x & - \frac{1}{4} v \\
0 & - D_x^2
\end{pmatrix}
, \qquad
L =
\begin{pmatrix}
- 2 & 0 \\
0 & 2 D_x
\end{pmatrix}
, \notag \\
K =
\begin{pmatrix}
0 & 0 \\
0 & - 1
\end{pmatrix}
, \qquad
s_i =
\begin{pmatrix}
p_i \\
r_i
\end{pmatrix}
. \label{e38}
\end{gather}
It is easy to find from \eqref{e36}--\eqref{e38} that
\begin{equation}
s_i = N_i s_{i+1} , \qquad i = 0 , 1 , \dotsc , n-3 , \label{e39}
\end{equation}
where
\begin{equation}
N_i =
\begin{pmatrix}
\frac{1}{2} D_x & - \frac{1}{8} v \\
0 & \frac{i + 1}{i + 2} D_x
\end{pmatrix}
, \label{e40}
\end{equation}
and to prove this by induction.

On the other hand,  the terms in \eqref{e26} with $\lambda^k$, for $k = n+2 , n+1 , n , n-1$, give us the following expressions:
\begin{gather}
r_n =0 , \qquad  p_n = 0 , \qquad r_{n-1} = - \tau_n v , \qquad p_{n-1} = \sigma_n u , \notag\\
r_{n-2} = - \tau_n v_x + ( \sigma_n - \tau_{n-1} ) v , \qquad p_{n-2} = \frac{1}{2} \sigma_n u_x + \frac{1}{8} \tau_n v^2  , \notag \\
r_{n-3} = - \tau_n v_{xx} + ( 2 \sigma_n - \tau_{n-1} ) v_x . \label{e41}
\end{gather}
It follows from \eqref{e39} and \eqref{e40} that
\begin{equation}
r_{n-3} = \frac{n-2}{n-1} D_x r_{n-2} , \label{e42}
\end{equation}
and this relation is satisfied by the expressions for $r_{n-2}$ and $r_{n-3}$ from \eqref{e41} if, and only if,
\begin{equation}
\tau_n = 0 , \qquad \tau_{n-1} = n \sigma_n . \label{e43}
\end{equation}
We take $\sigma_n = 1$ without loss of generality, and get from \eqref{e41} and \eqref{e43} that
\begin{equation}
s_n =
\begin{pmatrix}
0 \\
0
\end{pmatrix}
, \qquad
s_{n-1} =
\begin{pmatrix}
u \\
0
\end{pmatrix}
, \qquad
s_{n-2} =
\begin{pmatrix}
\frac{1}{2} u_x \\
( 1 - n ) v
\end{pmatrix}
.
\label{e44}
\end{equation}

All the other $s_i$, with $i = n-3 , n-4 , \dotsc , 1 , 0$, can be recursively obtained from $s_{n-2}$ \eqref{e44} via the relation \eqref{e39}, and we find from \eqref{e35} that
\begin{equation}
h^{(n)} = M \left( \prod_{i=0}^{n-3} N_i \right)
\begin{pmatrix}
\frac{1}{2} u_x \\
( 1 - n ) v
\end{pmatrix}
, \qquad n \ge 3 , \label{e45}
\end{equation}
where the matrix operators $M$ and $N_i$ are given by \eqref{e38} and \eqref{e40}. The order $n$ is not limited, all the expressions $h^{(n)}$ given by \eqref{e45} are local, the lower-order $h^{(0)}$, $h^{(1)}$ and $h^{(2)}$ are seen from \eqref{e29}--\eqref{e31}, and we have already got everything to express the matrices $T$ \eqref{e23} if necessary.

Consequently, the hierarchy, which the Ibragimov--Shabat system \eqref{e1} belongs to, is infinite and Lax-integrable.

\section{No recursion operator} \label{s5}

In our experience, a recursion operator can appear in a natural way, as a by-product, when an infinite integrable hierarchy is constructed by the cyclic basis method. In most of the studied cases, closure equations of cyclic bases lead to some recurrent relations, similar to \eqref{e35}--\eqref{e37} but with $K = 0$, and then recursion operators of the form $M L^{-1}$ appear straightforwardly \cite{S95,S04,S03a,SS,S03}. In a more complicated case \cite{KKS}, with $K \ne 0$, a recursion operator appears as well, owing to the specific condition $K L^{-1} K = 0$ satisfied there. This condition, however, is not satisfied by the matrices \eqref{e38}, and the relations \eqref{e39} do not lead to a recursion operator because the operator $N_i$ \eqref{e40} depends on $i$ in the present case. Since we have got the explicit expressions \eqref{e45} for the hierarchy members, we can directly investigate what is a recursion operator of this hierarchy, if there is any at all.

Let us represent the expressions \eqref{e45} in a more transparent form. We insert the unit matrix operator $I = M^{-1} M$ after every factor $N_i$ in \eqref{e45}, introduce the operator $J_i = M N_i M^{-1}$,
\begin{equation}
J_i =
\begin{pmatrix}
\frac{1}{2} D_x & \frac{i + 1}{4 ( i + 2 )} v D_x^{-1} \\
0 & \frac{i + 1}{i + 2} D_x
\end{pmatrix}
, \label{e46}
\end{equation}
use the identity
\begin{equation}
M
\begin{pmatrix}
\frac{1}{2} u_x \\
( 1 - n ) v
\end{pmatrix}
= J_{n-2}
\begin{pmatrix}
u_x \\
n v_x
\end{pmatrix}
\label{e47}
\end{equation}
with $J_{n-2}$ formally defined by \eqref{e46}, and obtain
\begin{equation}
h^{(n)} = \left( \prod_{i=0}^{n-2} J_i \right)
\begin{pmatrix}
u_x \\
n v_x
\end{pmatrix}
. \label{e48}
\end{equation}
This representation \eqref{e48} works also for $n = 2$ and extends \eqref{e45} in this sense. Moreover, if we insert the unit matrix
\begin{equation}
I =
\begin{pmatrix}
1 & 0 \\
0 & i + 2
\end{pmatrix}
\begin{pmatrix}
1 & 0 \\
0 & \frac{1}{i + 2}
\end{pmatrix}
\label{e49}
\end{equation}
after every factor $J_i$ in \eqref{e48}, we obtain
\begin{equation}
h^{(n)} = \left( \prod_{i=0}^{n-2} R_i \right)
\begin{pmatrix}
u_x \\
v_x
\end{pmatrix}
, \qquad n \ge 2 , \label{e50}
\end{equation}
where
\begin{equation}
R_i =
\begin{pmatrix}
1 & 0 \\
0 & \frac{1}{i + 1}
\end{pmatrix}
J_i
\begin{pmatrix}
1 & 0 \\
0 & i + 2
\end{pmatrix}
, \label{e51}
\end{equation}
that is,
\begin{equation}
R_i =
\begin{pmatrix}
\frac{1}{2} D_x & \frac{i + 1}{4} v D_x^{-1} \\
0 & D_x
\end{pmatrix}
. \label{e52}
\end{equation}

We have obtained the sufficiently simple representation \eqref{e50} for the hierarchy members, that is,
\begin{equation}
h^{(n)} = R_0 R_1 \dotsb R_{n-2} h^{(1)} , \qquad n \ge 2 , \label{e53}
\end{equation}
where $R_i$ are given by \eqref{e52}, and $h^{(1)}$ is, of course, the column of $u_x$ and $v_x$. Now, let us see whether the hierarchy members can be produced from each other by integer degrees of a single recursion operator. It follows from \eqref{e53} with \eqref{e52} that
\begin{equation}
h^{(2)} = R^{(2,1)} h^{(1)} , \qquad R^{(2,1)} = R_0 =
\begin{pmatrix}
\frac{1}{2} D_x & \frac{1}{4} v D_x^{-1} \\
0 & D_x
\end{pmatrix}
; \label{e54}
\end{equation}
that
\begin{gather}
h^{(3)} = R^{(3,2)} h^{(2)} , \notag \\
R^{(3,2)} = R_0 R_1 R_0^{-1} =
\begin{pmatrix}
\frac{1}{2} D_x & \frac{3}{8} v D_x^{-1} + \frac{1}{8} v_x D_x^{-2} \\
0 & D_x
\end{pmatrix}
; \label{e55}
\end{gather}
that
\begin{gather}
h^{(4)} = R^{(4,3)} h^{(3)} , \notag \\
R^{(4,3)} = R_0 R_1 R_2 R_1^{-1} R_0^{-1} \notag \\
\qquad \qquad =
\begin{pmatrix}
\frac{1}{2} D_x & - \frac{1}{16} v D_x^{-1} + \frac{1}{16} v_{xx} D_x^{-3} \\
0 & D_x
\end{pmatrix}
; \label{e56}
\end{gather}
and so on. The operators $R^{(2,1)}$ \eqref{e54}, $R^{(3,2)}$ \eqref{e55} and $R^{(4,3)}$ \eqref{e56} are not equal to each other. No surprise, because the condition $R^{(k+3,k+2)} = R^{(k+2,k+1)}$, satisfied for some $k$, would lead to the condition $R_{k+1} = R_k$ which is impossible due to \eqref{e52}.

Consequently, the hierarchy, which the Ibragimov--Shabat system \eqref{e1} belongs to, does not possess a recursion operator.

\section{Conclusion} \label{s6}

We studied the Lax integrability of one nonlinear evolutionary system, explicitly constructed its infinite integrable hierarchy, and discovered that the hierarchy has no recursion operator.

It would also be interesting to study the Lax integrability and hierarchies of some other Bakirov-type systems known to possess only finite numbers of local generalized symmetries.


\begin{thebibliography}{99}

\small

\bibitem{IS} N.K. Ibragimov, A.B. Shabat, Evolutionary equations with nontrivial Lie--B\"{a}cklund group, Funct. Anal. Its Appl. 14 (1980) 19--28.

\bibitem{Olv} P.J. Olver, Applications of Lie Groups to Differential Equations, Springer, New York, 1986 (1st ed.) and 1993 (2nd ed.).

\bibitem{BSW} F. Beukers, J.A. Sanders, J.P. Wang, On integrability of systems of evolution equations, J. Diff. Eq. 172 (2001) 396--408.

\bibitem{Ser} A. Sergyeyev, On a class of inhomogeneous extensions for integrable evolution systems, In: Differential Geometry and Its Applications (Proc. 8th Conf.), Silesian Univ., Opava, 2001, pp. 243--252; arXiv:nlin/0310032.

\bibitem{S95} S.Yu. Sakovich, On zero-curvature representations of evolution equations, J. Phys. A: Math. Gen. 28 (1995) 2861--2869.

\bibitem{S04} S.Yu. Sakovich, Cyclic bases of zero-curvature representations: five illustrations to one concept, Acta Appl. Math. 83 (2004) 69--83; arXiv:nlin/0212019.

\bibitem{S03a} S.Yu. Sakovich, Cyclic bases of zero-curvature representations: further examples, arXiv:nlin/0311027.

\bibitem{S99} S.Yu. Sakovich, Coupled KdV equations of Hirota--Satsuma type, J. Nonlinear Math. Phys. 6 (1999) 255--262; arXiv:solv-int/9901005.

\bibitem{KSY} A. Karasu-Kalkanl\i, S.Yu. Sakovich, \'{I}. Yurdu\c{s}en, Integrability of Kersten--Krasil'shchik coupled KdV-mKdV equations: singularity analysis and Lax pair, J. Math. Phys. 44 (2003) 1703--1708; arXiv:nlin/0206046.

\bibitem{S11} S. Sakovich, Integrability of the Bakirov system: a zero-curvature representation, Int. J. Math. Math. Sci. 2011 (2011) 497828; arXiv:nlin/0206034.

\bibitem{SS} A. Sakovich, S. Sakovich, The short pulse equation is integrable, J. Phys. Soc. Japan 74 (2005) 239--241; arXiv:nlin/0409034.

\bibitem{S18} S. Sakovich, On a new avatar of the sine-Gordon equation, Nonlinear Phenom. Complex Syst. 21 (2018) 62--68; arXiv:1703.04678.

\bibitem{Mar} M. Marvan, On zero-curvature representations of partial differential equations, In: Differential Geometry and Its Applications (Proc. 5th Conf.), Silesian Univ., Opava, 1993, pp. 103--122.

\bibitem{S94} S.Yu. Sakovich, On conservation laws and zero-curvature representations of the Liouville equation, J. Phys. A: Math. Gen. 27 (1994) L125--L129.

\bibitem{S20} S. Sakovich, True and fake Lax pairs: how to distinguish them, Nonlinear Phenom. Complex Syst. 23 (2020) 338--341; arXiv:nlin/0112027.

\bibitem{S03} S.Yu. Sakovich, On integrability of one third-order nonlinear evolution equation, Phys. Lett. A 314 (2003) 232--238; arXiv:nlin/0303040.

\bibitem{S14} S. Sakovich, A note on Lax pairs of the Sawada--Kotera equation, J. Math. 2014 (2014) 906165; arXiv:1402.0127.

\bibitem{S05a} S. Sakovich, On continual classes of evolution equations, arXiv:nlin/0501038.

\bibitem{S05} S. Sakovich, Enlarged spectral problems and nonintegrability, Phys. Lett. A 345 (2005) 63--68; arXiv:nlin/0504037.

\bibitem{KKS} A. Karasu-Kalkanl\i, A. Karasu, S.Yu. Sakovich, A strange recursion operator for a new integrable system of coupled Korteweg--de~Vries equations, Acta Appl. Math. 83 (2004) 85--94; arXiv:nlin/0203036.

\end{thebibliography}
\end{document}